\documentclass[onecolumn,showpacs,preprintnumbers,amsmath,amssymb]{revtex4}
\usepackage{graphicx,epsf}

\newcommand{\beq}{\begin{eqnarray}}
\newcommand{\eeq}{\end{eqnarray}}
\newcommand{\eps}{\varepsilon}
\newcommand{\btem}{\bibitem}

\def\Vec#1{\mbox{\boldmath$#1$}}
\def\Ren#1{\mbox{\boldmath$#1$}}

\begin{document}
 \draft

\title{Derivation of Covariant Dissipative Fluid Dynamics
in the Renormalization-group Method}

\author{K. Tsumura}
\affiliation{Department of Physics,
 Kyoto University, Kyoto 606-8502, Japan}

\author{T. Kunihiro}
\affiliation{Yukawa Institute for Theoretical Physics,
Kyoto University, Kyoto 606-8502, Japan}

\author{K. Ohnishi}
\affiliation{Yukawa Institute for Theoretical Physics,
Kyoto University, Kyoto 606-8502, Japan}

\begin{abstract}
We derive generic relativistic
 hydrodynamical equations with dissipative effects
 from the underlying Boltzmann equation in a mechanical and systematic
 way on the basis of so called the renormalization-group
method. A macroscopic frame vector is introduced to specify the
frame on which  the macroscopic dynamics is described.
Our method is so mechanical with only few
ansatz that our method give a microscopic foundation of the
available hydrodynamical equations, and also can be applied to
make a reduction of the kinetic equations other than the simple
Boltzmann equation.
\end{abstract}

\date{\today}
\maketitle

The relativistic hydrodynamical equation is widely used
in various fields of physics, especially in
 high-energy nuclear physics and astrophysics
\cite{kolb,Gyulassy:2004zy,dis001}, and it seems that
the study of the {\em dissipative} hydrodynamical equations
is now becoming a central interest in these fields
\cite{dis001,dis002,dis003,dis004,dis005,dis006,nonaka06,ast001}.

 It is, however, noteworthy
that  we have not necessarily reached a full understanding of the
 theory of relativistic hydrodynamics for viscous fluids:
(1) It is well known that
the relativistic hydrodynamical equation with dissipation
has a form  depending on the choice of the Lorentz frame or
the definition of the hydrodynamical flow.
The typical frame is that on the particle or the energy flow,
and the typical corresponding equations are the ones of
Eckart \cite{hen001} and Landau \cite{hen002},
which were derived phenomenologically
on the basis of the number and energy-momentum conservation laws, the
second law of thermodynamics and some specific assumptions on the choice
of the flow. One might consider that
the difference of the equations should
 be merely a kind of choice of coordinate
to describe dynamics in a easy and practical way, and
one might be able to go back and forth between the two reference frames,
by a Lorentz transformation.
However, it may not be the case.
To be more specific, let
 $\delta T^{\mu\nu}$ and $\delta N^\mu$,
be the dissipative part of the energy-momentum tensor and the particle
 current, respectively. The point is that they are not determined uniquely
  only by the second law of thermodynamics without some physical
 constraints involving the four flow-vector $u_{\mu}$ with
$u_{\mu}u^{\mu}=1$:
  The constraints of Eckart are expressed as \cite{cau001},\,
(i)\, $u_\mu  u_\nu  \delta T^{\mu\nu} = 0$,\,
(ii)\, $u_\mu  \delta N^\mu = 0$,\,
(iii)\, $\Delta_{\mu\nu} \delta N^\nu = 0$,
 while those of Landau are \,
(i)\, $u_\mu  u_\nu  \delta T^{\mu\nu} = 0$,\,
(ii)\, $u_\mu  \delta N^\mu = 0$\, and
(iv)\, $u_\mu  \Delta_{\nu\rho} \delta T^{\mu\nu} = 0$,\,
where $\Delta_{\mu\nu}=g_{\mu\nu}-u_{\mu}u_{\nu}$.
As one sees, the first two constraints are the same for the two
theories, and the respective third condition specifies the frame of the
coordinate.
It is here  noteworthy that there is a proposal by
Stewart \cite{mic003} for
the constraints for the Eckart frame, i.e., the particle-flow frame,
as given by
(v)\, $\delta T^{\mu}_{\,\,\,\mu}=0$,\,
(ii)\, $u_\mu  \delta N^\mu = 0$,\,
(iii)\, $\Delta_{\mu\nu} \delta N^\nu = 0$,
where the condition (i) of Eckart is replaced by a different one (v).
One may ask if both the Eckart and Stewart constraints make sense or not.
It has been scarcely examined \cite{vanweert}
whether the constraints proposed
phenomenologically hold internal consistency or match the underlying
 microscopic kinetic theories.
(2) It is also known that these equations containing the time derivative
only in the first order, unfortunately, suffer from
the unphysical instability \cite{cau001,cau002};
the instability is attributed to the lack of the causality,
 and invented is a
fluid dynamical equation which contains the second-order time-derivative
 as well as the first-order one \cite{hen001,hen002,hen003}.
The new equation called Israel-Stewart equation \cite{mic004}
contains new transport
coefficients like the relaxation time yet to be determined
microscopically or phenomenologically \cite{mic001,mic002,mic003,mic004}.

There are some attempts to
 derive the phenomenological equations from the microscopic
equation, i.e., the relativistic Boltzmann equation,
 with use of the Chapman-Enskog method \cite{chapman}
 and the Grad's fourteen-moment method \cite{grad}, and hence
give a microscopic foundation to them
\cite{mic001,mic002,mic003,mic004,mic005}.
We notice that such a microscopic approach has another merit
that the hydrodynamical equations can be obtained, which are
consistent with the underlying kinetic
 equation,
because the theory gives the theory connect the
kinetic and the hydrodynamical regimes and hence
 gives the natural initial condition for the
hydrodynamical equation. This point
should have a practical merit when one analyzes the
freeze-out regime and also the hadron corona in the expanding
system such as the intermediate stage of the relativistic heavy-ion
collision \cite{dis005,nonaka06,connect}.

One must say, however, that the past works in the
microscopic approaches are not fully satisfactory in giving the
foundation to the phenomenological equations:
{\bf a.}~Although the past works succeeded in
identifying the assumptions and/or approximations
  to reproduce the known hydrodynamical equations
by Eckart, Landau and Israel,
one must say that
the physical meaning of these assumptions/approximations is somewhat
obscure,
and so is the uniqueness of
those hydrodynamical equations as
 the long-wavelength and
 low-frequency limit of the microscopic theory.
{\bf b.}~It is also to be noted that
  the Chapman-Enskog method and the Grad's
 fourteen-moment method themselves have some ad-hoc part and may not
be fully mechanical and systematic methods for the reduction
of dynamics.

 The purpose of this article is to derive the relativistic
 hydrodynamical equations with dissipative effects
 from the microscopic theory in a more natural and systematic way,
 and thereby establish the microscopic foundation of them.
We shall show that when the particle-flow frame (Eckart frame) is
chosen, the constraints by Stewart but not Eckart must be taken,
while the resulting hydrodynamical equation which  manifestly
satisfies the second law of thermodynamics is different form
that given by Stewart.

The problem is how to reduce the dynamics to a slower dynamics
using a reliable reduction theory of the dynamics \cite{kuramoto}, which
should be as mechanical as possible.
In fact,
van Kampen \cite{mic005} applied his method \cite{kampen2} of the reduction
to derive the
relativistic hydrodynamic equations for the viscous fluid.
However, his method is not formulated in a covariant way
 and hence failed in reproducing the
phenomenological hydrodynamical equations with dissipation.
We shall apply another powerful reduction theory known as
the ``renormalization-group(RG) method''
 \cite{rgm001,env001,env002,env005,env006,env007,env008}
to reduce the relativistic Boltzmann equation to the
 hydrodynamical equations.
Our theory is manifestly covariant and relies on almost no assumptions.

The RG method is a systematic reduction theory of the dynamics
leading to the coarse-graining of temporal and spatial scales.
This method is suitable for our present purpose to
elucidate the physical meanings of each process
of the reduction since the method is so mechanical
and  does not  require any specific assumptions.
Indeed, the RG method is already applied satisfactorily
to the derivation of the
Navier-Stokes equation from the
(non-relativistic) Boltzmann equation \cite{env007,env008}.
The present work is an attempt to extend these previous
works to the relativistic case.


In this article, we shall treat, as a simplest example, the
 classical but relativistic system composed of
 identical particles; an extension to multi-component systems is
possible and will be discussed in a separate paper \cite{extend}.
Then the relativistic Boltzmann equation \cite{mic001} describing such a
system
reads
 \begin{eqnarray}
  \label{eq:001}
   p^\mu \, \partial_\mu f_p(x) = C[f]_p(x),
 \end{eqnarray}
 where
$f_p(x)$ denotes the one-particle distribution function
 defined in the phase space $(x^{\mu} \,,\, p^{\mu})$
with $p^\mu$ being the four momentum of the on-shell particle;
 $p^\mu \, p_\mu \equiv p^2 = m^2$ and $p^0 \ge 0$.
 The right-hand side of Eq.(\ref{eq:001}) is called the collision integral,
$C[f]_p(x) \equiv \frac{1}{2!} \,
   \sum_{p_2} \, \frac{1}{p_2^0} \, \sum_{p_3} \,
   \frac{1}{p_3^0} \, \omega(p \,,\, p_1|p_2 \,,\, p_3) \,
   \big( f_{p_2}(x) \, f_{p_3}(x) - f_p(x) \, f_{p_1}(x) \big),
$
 where $\omega(p \,,\, p_1|p_2 \,,\, p_3)$ denotes
 the transition probability owing to the microscopic two-particle
 interaction
and  contains the delta functions $\delta^4(p + p_1 - p_2 - p_3)$
representing
 the energy-momentum conservation.
The transition probability has the symmetric properties due
to the indistinguishability
 of the particles and the time reversal invariance in the scattering
 process:
 $\omega(p \,,\, p_1|p_2 \,,\, p_3) = \omega(p_2 \,,\, p_3|p \,,\, p_1)
 = \omega(p_1 \,,\, p|p_3 \,,\, p_2) = \omega(p_3 \,,\, p_2|p_1 \,,\, p)$.
It should be stressed here that we have
 confined ourselves to the case in which
the particle number is conserved in the collision process.

The particle-number and energy-momentum conservation in the
collision process ensures the following important relations;
$\sum_p \, \frac{1}{p^0} \, C[f]_p(x) = \sum_p \, \frac{1}{p^0} \, p^\mu \,
C[f]_p(x) = 0$,
which says that the five quantities $(1 \,,\, p^\mu)$ are {\em collision
invariants}.
We have the continuity or balance equations
for the particle current
and the energy-momentum tensor,
\begin{eqnarray}
  \label{eq:balance}
  \partial_\mu \sum_p \, \frac{1}{p^0} \, p^\mu \, f_p(x) \equiv
  \partial_\mu N^\mu =0, \,\,\,\,
  \partial_\nu \sum_p \, \frac{1}{p^0} \, p^\mu \, p^\nu \, f_p(x)
  \equiv \partial_\nu T^{\mu\nu} = 0,
\end{eqnarray}
respectively.
Notice that while these equations have the same forms as the hydrodynamical
equations, nothing about the dynamical properties
is contained in these equations
before the evolution of the distribution function
$f_p(x)$ is obtained by solving the microscopic Eq.(\ref{eq:001}).

The entropy current may be defined by
$S^\mu \equiv - \sum_p \, \frac{1}{p^0} \, p^\mu \, f_p(x) \, (\ln f_p(x) -
1)$,
which is conserved
only if $\ln f_p(x)$ is a collision invariant,
or a linear combination of the basic collision invariants
$(1 \,,\, p^\mu)$
as $\ln f_p(x) = \alpha(x) + p^\mu \, \beta_\mu(x)$
with $\alpha(x)$ and $\beta_\mu(x)$ being arbitrary functions.

Since we are interested in the hydrodynamical regime where
the time- and  space-dependence of the physical
quantities are small,
we solve Eq.(\ref{eq:001}) under the situation where the
time variation of $f_p(x)$ is slow with long wavelengths.

To retain the Lorentz covariance,
and achieve a coarse-graining,
it is found convenient to introduce a
macroscopic
Lorentz vector $\Ren{a}_p^\mu$
characterizing the flow, which
specifies the covariant coordinate system (frame) and we call
the {\em macroscopic-frame vector}.
Although $\Ren{a}_p^\mu$ may depend on the momentum $p$ and the space-time
coordinate $x$, the time variation of it is supposed to be
much smaller than that of the microscopic processes. Our point is
that the separation of the scales can be nicely achieved by the RG method.

Keeping in mind the above scale difference,
we define the covariant
coordinate system $(\tau \,,\, \sigma^\mu)$ as
$\mathrm{d}\tau \equiv \Ren{a}_p^\mu(x) \, \mathrm{d}x_\mu$
and   $\varepsilon^{\mbox{-}1} \, \mathrm{d}\sigma^\mu
  \equiv \big(
  g^{\mu\nu} -
\Ren{a}_p^\mu(x) \, \Ren{a}_p^\nu(x) / \Ren{a}_p^2(x) \big) \,
\mathrm{d}x_\nu
  \equiv \Ren{\Delta}_p^{\mu\nu}(x) \, \mathrm{d}x_\nu$.
We notice that  the small quantity
$\varepsilon$ have been  introduced to tag that the space derivatives
are small for the system we are interested in;
the $\varepsilon$ may be identified with the ratio of the average
particle distance over the mean free path.
In this coordinate system, Eq.(\ref{eq:001}) reads
\begin{eqnarray}
  \label{eq:004}
  \frac{\partial}{\partial \tau} f_p(\tau \,,\, \sigma) =
  \frac{1}{p \cdot \Ren{a}_p(\tau \,,\, \sigma)} \,
  C[f]_p(\tau \,,\, \sigma)
  - \varepsilon \, \frac{1}{p \cdot \Ren{a}_p(\tau \,,\, \sigma)} \, p \cdot
  \Ren{\nabla} f_p(\tau \,,\, \sigma),
\end{eqnarray}
where $\Ren{a}_p^\mu(\tau \,,\, \sigma) \equiv \Ren{a}_p^\mu(x)$,
$\Ren{\Delta}_p^{\mu\nu}(\tau \,,\, \sigma) \equiv
\Ren{\Delta}_p^{\mu\nu}(x)$
and $f_p(\tau \,,\, \sigma) \equiv f_p(x)$.
Since $\varepsilon$ appears in front of
$\Ren{\nabla}^\mu \equiv \Ren{\Delta}_p^{\mu\nu}(\tau \,,\, \sigma) \,
\frac{\partial}{\partial \sigma^\nu}$,
Eq.(\ref{eq:004}) has a form to which the perturbative expansion
can be possible.
In fact,   a natural but significant assumption
underlies this seemingly mere rewrite of the equation that
 only the spatial inhomogeneity over distances of the
order of the mean free path is the origin of the dissipation.
We notice that the RG method
applied to non-relativistic Boltzmann equation
with the corresponding assumption nicely leads to the Navier-Stokes
equation \cite{env007,env008};
the present approach is simply a covariantization of the non-relativistic
case.

Some remarks on $\Ren{a}_p^\mu(\tau \,,\, \sigma)$ are in order:
(i) In the perturbative expansion,
we shall take the coordinate system
where
$\Ren{a}_p^\mu(\tau \,,\, \sigma)$
has no $\tau$ dependence, i.e.,
$\Ren{a}_p^\mu(\tau \,,\, \sigma)= \Ren{a}_p^\mu(\sigma)$.
(ii) In our coordinate system,
the physical quantity transported
per unit $\tau$ is given by $p \cdot \Ren{a}_p(\sigma)$,
we can control what is the flow represented in
our theory by varying the specific expression of
$\Ren{a}_p^\mu(\sigma)$.
Owing to this freedom inherent in our coordinate system
our theory may lead
to various hydrodynamical equations,
including the ones in the  Eckart and Landau frames
for non-ideal fluids.

In accordance with the general formulation of the RG method
given in \cite{env001,env002,env006},
we first try to obtain the perturbative
solution $\tilde{f}_p$ around
the arbitrary initial time $\tau = \tau_0$ with the initial value
$f_p(\tau_0 \,,\, \sigma)$;
$\tilde{f}_p(\tau=\tau_0 \,,\, \sigma \,;\, \tau_0)=f_p(\tau_0 \,,\,
\sigma)$,
where we have made explicit that the solution
have the $\tau_0$ dependence.
The initial value is not yet specified,
we suppose that the initial value is on
an exact solution with $\tau_0$ being varied.
The initial value as well as the solution are expanded
with respect to $\varepsilon$ as follows;
$\tilde{f}_p(\tau \,,\, \sigma \,;\, \tau_0) = \tilde{f}_p^{(0)}(\tau
\,,\, \sigma \,;\, \tau_0) + \varepsilon \, \tilde{f}_p^{(1)}(\tau
\,,\, \sigma \,;\, \tau_0) + \varepsilon^2 \, \tilde{f}_p^{(2)}(\tau
\,,\, \sigma \,;\, \tau_0) + \cdots$,
and
$f_p(\tau_0 \,,\, \sigma) = f_p^{(0)}(\tau_0 \,,\, \sigma) +
\varepsilon \, f_p^{(1)}(\tau_0 \,,\, \sigma) + \varepsilon^2 \,
f_p^{(2)}(\tau_0 \,,\, \sigma) + \cdots$.
The respective initial conditions at $\tau = \tau_0$ are set up as
$\tilde{f}_p^{(l)}(\tau_0 \,,\, \sigma \,;\, \tau_0)
= f_p^{(l)}(\tau_0 \,,\, \sigma) \,\,\, \mbox{for} \,\,\, l = 0, \, 1, \,
2,\cdots$.
In the expansion, the zeroth-order value
$\tilde{f}_p^{(0)}(\tau_0 \,,\, \sigma \,;\, \tau_0) = f_p^{(0)}(\tau_0
\,,\, \sigma)$
is supposed to be as close as possible
to an exact solution.

Substituting the above expansions into Eq.(\ref{eq:004})
in the $\tau$-independent but $\tau_0$-dependent coordinate system
with $\Ren{a}^\mu_p(\tau \,,\, \sigma) \rightarrow \Ren{a}^\mu_p(\tau_0
\,,\, \sigma)
\equiv \Ren{a}^\mu_p(\sigma \,;\, \tau_0)$,
we obtain the
series of the perturbative equations with respect to $\varepsilon$.
The zeroth-order equation
reads
\begin{eqnarray}
  \label{eq:1st}
  \frac{\partial}{\partial \tau} \tilde{f}^{(0)}_p
  = \frac{1}{p \cdot \Ren{a}_p} \, C[\tilde{f}^{(0)}]_p.
\end{eqnarray}
Since we are interested in the slow motion which may be realized
asymptotically when $\tau \rightarrow \infty$,
we should take the following stationary solution or the fixed
point,
$\frac{\partial}{\partial \tau}\tilde{f}_p^{(0)} = 0$,
which is realized when
$C[\tilde{f}^{(0)}]_p = 0$
for arbitrary $\sigma$.
This implies
that
$\ln \tilde{f}_p^{(0)}$ is a linear combination of
the five collision invariants $(1 \,,\, p^\mu)$, and hence
$\tilde{f}_p^{(0)}$ is found to be
a local equilibrium distribution function or
the Juettner function (the relativistic analog of the Maxwellian):
$\tilde{f}_p^{(0)}(\tau \,,\, \sigma \,;\, \tau_0) =
  (2\pi)^{\mbox{-}3} \exp \big[
    (\mu(\sigma \,;\, \tau_0) - p^\mu \, u_\mu(\sigma \,;\, \tau_0)) \, / \,
     T(\sigma \,;\, \tau_0) \big]$
  $\equiv f^{\mathrm{eq}}_p(\sigma \,;\, \tau_0)$
with $u^\mu(\sigma \,;\, \tau_0) \, u_\mu(\sigma \,;\, \tau_0) = 1$,
which implies
that
$\tilde{f}_p^{(0)}(\tau \,,\, \sigma \,;\, \tau_0)
= \tilde{f}_p^{(0)}(\tau_0 \,,\, \sigma \,;\, \tau_0)
= f_p^{(0)}(\tau_0 \,,\, \sigma)
= f^{\mathrm{eq}}_p(\sigma \,;\, \tau_0)$.
It should be noticed that the five would-be integration constants
 $T(\sigma \,;\, \tau_0)$, $\mu(\sigma \,;\, \tau_0)$ and
 $u_\mu(\sigma \,;\, \tau_0)$
is independent of $\tau$ but may depend on $\tau_0$ as well as
$\sigma$.

Next the first-order equation reads
 \begin{eqnarray}
  \label{eq:006}
  \frac{\partial}{\partial \tau} \tilde{f}_p^{(1)} = \sum_q \, A_{pq} \,
\tilde{f}_q^{(1)} + F_p,
 \end{eqnarray}
 where the linear evolution operator $A$ and the inhomogeneous term
 $F$ are defined as
$A_{pq} \equiv (p \cdot \Ren{a}_p)^{\mbox{-}1} \, \frac{\partial}{\partial
f_q}
  C[f]_p \big|_{f = f^{\mathrm{eq}}}$ and
  $F_p \equiv - (p \cdot \Ren{a}_p)^{\mbox{-}1} \, p \cdot \Ren{\nabla}
f^{\mathrm{eq}}_p$,
 respectively.
 To obtain the solution which describes the slow motion,
 it is convenient to first analyze the properties of $A$,
 especially its spectra.
 For this purpose, we convert $A$ to
 another linear evolution operator
 $L \equiv f^{\mathrm{eq}\mbox{-}1} \, A \, f^{\mathrm{eq}}$ with
 the diagonal matrix
 $f^\mathrm{eq}_{pq} \equiv f^\mathrm{eq}_p \, \delta_{pq}$.
 Let us define the inner product between arbitrary vectors $\varphi$
 and $\psi$ by
\begin{eqnarray}
\langle  \, \varphi \,,\, \psi \, \rangle
   \equiv \sum_{p} \, \frac{1}{p^0} \, (p \cdot \Ren{a}_p) \,
f^{\mathrm{eq}}_p \,
\varphi_p \, \psi_p.
\label{eq:inner}
\end{eqnarray}
 With this inner product,
 one can nicely show that $L$ becomes self-adjoint
 $\langle \, \varphi \,,\, L \, \psi \, \rangle = \langle \, L \, \varphi
\,,\, \psi \, \rangle$
 and non-positive definite
 $\langle \, \varphi \,,\, L \, \varphi \, \rangle \le 0$, which
 means that the eigen values of $L$ are zero or negative.
 The eigen vectors of the zero eigen value
 are
$\varphi_{0p}^\alpha = p^\mu$ for
$\alpha = \mu = 0 \sim 3$ and
$\varphi_{0p}^4 = m$,
 which span the kernel of $L$ and satisfy $L \, \varphi_{0}^\alpha = 0$.
 Notice that these five vectors are the collision invariants.
 Following \cite{env006}, we define the projection operator $P$
 onto the kernel of $L$ which is
 called the P space and the projection operator $Q$ onto the Q space
 complement to the P space:
\begin{eqnarray}
  \big[ P \, \psi \big]_p \equiv \sum_{\alpha\beta} \, \varphi_{0p}^\alpha
\, \eta^{\mbox{-}1}_{\alpha\beta} \,
  \langle \, \varphi_0^\beta \,,\, \psi \, \rangle\label{eq:010},
\end{eqnarray}
with $Q\equiv 1 - P$ where $\eta^{\mbox{-}1}_{\alpha\beta}$
 is the inverse matrix of
 $\eta^{\alpha\beta} \equiv \langle \, \varphi_0^\alpha \,,\,
\varphi_0^\beta \, \rangle$.

 In the following,
 we shall suppress the argument $\sigma$
 and the subscript $p$ when no misunderstanding is expected.
 The solution to Eq.(\ref{eq:006}) with the initial condition
 $\tilde{f}^{(1)}(\tau = \tau_0 \,;\, \tau_0) = f^{(1)}(\tau_0)$
 is found to be
 $\tilde{f}^{(1)}(\tau \,;\, \tau_0) = \mathrm{e}^{(\tau - \tau_0)A}
 \big\{ f^{(1)}(\tau_0) + A^{\mbox{-}1} \, \bar{Q} \, F \big\}
 + (\tau - \tau_0) \, \bar{P} \, F - A^{\mbox{-}1} \, \bar{Q} \, F$,
 where the modified projection operators
 $\bar{P} \equiv f^{\mathrm{eq}} \, P \, f^{\mathrm{eq}\mbox{-}1}$ and
 $\bar{Q} \equiv f^{\mathrm{eq}} \, Q \, f^{\mathrm{eq}\mbox{-}1}$
 have been introduced.
 Notice that the first term  would be a fast motion coming from the Q space,
 which we can nicely eliminate by choosing the initial value
 $f^{(1)}(\tau_0)$,  which has  not yet been specified;
 thus we have  the first-order solution
 \begin{eqnarray}
   \label{eq:1st-sol}
   \tilde{f}^{(1)}(\tau \,;\, \tau_0) =
   (\tau - \tau_0) \, \bar{P} \, F - A^{\mbox{-}1} \, \bar{Q} \, F
 \end{eqnarray}
 with the initial value determined as
 $\tilde{f}^{(1)}(\tau_0 \,;\, \tau_0)= f^{(1)}(\tau_0) = - A^{\mbox{-}1} \,
\bar{Q} \, F$.
 We notice the appearance of the
 secular term proportional to $\tau - \tau_0$,
 which apparently invalidates the perturbative solution
 when $\vert \tau - \tau_0 \vert$ becomes large.

 The second-order equation
 \footnote{
 Although the bilinear term of $\tilde{f}^{(1)}(\tau)$ appears from the
 collision integral, here we neglect it.
 It is known that the neglected term produces the so-called Burnett
 terms which represents the higher-order non-equilibrium effects.
 }
 is written as
 \begin{eqnarray}
  \frac{\partial}{\partial \tau} \tilde{f}^{(2)} = A
   \, \tilde{f}^{(2)} + (\tau - \tau_0) \, H + I,
 \end{eqnarray}
 with
  $H_p \equiv - (p \cdot \Ren{a}_p)^{\mbox{-}1} \, p \cdot \Ren{\nabla}
   [\bar{P} \, F ]_p$ and
  $I_p \equiv (p \cdot \Ren{a}_p)^{\mbox{-}1} \, p \cdot \Ren{\nabla}
   [ A^{\mbox{-}1} \, \bar{Q} \, F ]_p$,
the solution to which is found to be
 \begin{eqnarray}
  \label{eq:013}
 \tilde{f}^{(2)}(\tau \,;\, \tau_0) &= &\mathrm{e}^{(\tau - \tau_0)A} \,
\big\{ f^{(2)}(\tau_0)
   + A^{\mbox{-}2} \, \bar{Q} \, H + A^{\mbox{-}1} \, \bar{Q} \, I
   \big\}
\nonumber\\
  & &  + \frac{1}{2} \, (\tau - \tau_0)^2 \, \bar{P} \, F
   + (\tau - \tau_0) \, \big\{ \bar{P} \, I - A^{\mbox{-}1} \, \bar{Q}
   \, H \big\}
   - \big\{ A^{\mbox{-}2} \, \bar{Q} \, H + A^{\mbox{-}1} \, \bar{Q} \,
   I \big\}.
 \end{eqnarray}
Again the would-be fast motion can be eliminated
by the choice of the initial value $f^{(2)}(\tau_0)$,
 and thus we have the second-order solution
\begin{eqnarray}
 \label{eq:sol-2nd}
\tilde{f}^{(2)}(\tau \,;\, \tau_0) =
\frac{1}{2} \, (\tau - \tau_0)^2 \, \bar{P} \, F
 + (\tau - \tau_0) \, \big\{ \bar{P} \, I - A^{\mbox{-}1} \, \bar{Q} \,
  H \big\}
 - \big\{ A^{\mbox{-}2} \, \bar{Q} \, H + A^{\mbox{-}1} \, \bar{Q} \,
  I \big\},
\end{eqnarray}
with the initial value determined as
 $\tilde{f}^{(2)}(\tau_0 \,;\, \tau_0)=f^{(2)}(\tau_0) =
- A^{\mbox{-}2} \, \bar{Q} \, H - A^{\mbox{-}1} \, \bar{Q} \, I$.
 We notice again the appearance of the secular terms.

Summing up the perturbative solutions up to the second order,
we have an approximate solution around
 $\tau \simeq \tau_0$ to this order;
 $\tilde{f}(\tau \,;\, \tau_0) = \tilde{f}^{(0)}(\tau \,;\, \tau_0) +
\varepsilon \,
 \tilde{f}^{(1)}(\tau \,;\, \tau_0) + \varepsilon^2 \,
 \tilde{f}^{(2)}(\tau \,;\, \tau_0) + O(\varepsilon^3)$,
containing the secular terms  which apparently
 invalidates the perturbative expansion
 for $\tau$ away from the initial time $\tau_0$.

The point of the RG method lies in the fact that
 we can utilize the secular terms to obtain
a solution valid in a global domain.
Now we may see that we have a family of curves $\tilde{f}(\tau \,;\,
\tau_0)$
 parameterized with $\tau_0$.
 They are all on the exact solution $f(\tau)$ at $\tau = \tau_0$
 up to $O(\eps ^3)$, but
 only valid locally for $\tau$ near $\tau_0$.
 So it is conceivable that the envelope
 $E$ of the family of curves which
 contacts with each local solution at $\tau=\tau_0$ will
give a global solution in our asymptotic situation.
 According to the classical theory of envelopes,  the
envelope which contact with any
curve in the family at $\tau=\tau_0$ is obtained by
$ \frac{\mathrm{d}}{\mathrm{d}\tau_0} \tilde{f}_p(\tau \,,\, \sigma \,;\,
\tau_0)
   \big|_{\tau_0 = \tau} = 0$,
or explicitly
 \begin{eqnarray}
  \label{eq:015}
   \frac{\partial}{\partial \tau} \big\{ f_p^{\mathrm{eq}} -
   \varepsilon \, \big[ A^{\mbox{-}1} \, \bar{Q} \, F \big]_p \big\}
   - \varepsilon \, \big[ \bar{P} \, F \big]_p
   - \varepsilon^2 \, \big\{ \big[ \bar{P} \, I \big]_p - \big[
A^{\mbox{-}1} \, \bar{Q} \,
   H \big]_p \big\} + O(\varepsilon^3) = 0.
 \end{eqnarray}
 This envelope equation is the basic equation in the RG method
and gives the equation of motion governing the dynamics of
 the five slow variables $T(\tau)$, $\mu(\tau)$ and $u^\mu(\tau)$
 in $f_p^\mathrm{eq}$.
The global solution in the asymptotic region is
given as the envelope function,
\begin{eqnarray}
\label{eq:manifold}
 \tilde{f}_p(\tau \,,\, \sigma \,;\, \tau_0=\tau) \equiv f_p(\tau \,,\,
\sigma)
 = f_p^\mathrm{eq} - \varepsilon \big[ A^{\mbox{-}1} \, \bar{Q} \, F
\big]_p -
 \varepsilon^2 \big\{ \big[ A^{\mbox{-}2} \, \bar{Q} \, H \big]_p + \big[
A^{\mbox{-}1} \,
\bar{Q} \, I \big]_p \big\} + O(\varepsilon^3),
\end{eqnarray}
 where the exact solution of Eq.(\ref{eq:015}) is inserted.
Thus one sees that  $f_p(\tau)$ now describes
the macroscopic-time
 evolution of the one-particle distribution function in Eq.(\ref{eq:001}),
because the time-derivatives of the quantities in
$f_p(\tau)$  are all
in the order of $\varepsilon$.
We emphasize that we have derived
the infrared effective theory of Eq.(\ref{eq:001})
in the form of the pair of Eq.'s (\ref{eq:015}) and (\ref{eq:manifold}).
This is one of the main results in this article.

 Now the RG/Envelop equation Eq.(\ref{eq:015})
is actually the hydrodynamic equation governing
the the five slow variables $T(\tau)$, $\mu(\tau)$ and $u^\mu(\tau)$.
To show this explicitly,
 we apply $\bar{P}$ from the left and then take the inner product with
 the five zero modes $\varphi_0^\alpha$\footnote{This is tantamount to
inserting the solution $f(\tau)$ into Eq.(\ref{eq:balance}).}.
 Putting back $\varepsilon = 1$, we arrive at
$\partial_\mu J^{\mu\alpha}_{\mathrm{hydro}}= 0$ with
\beq
  \label{eq:017}
    J^{\mu\alpha}_{\mathrm{hydro}} &\equiv &\sum_{p} \, \frac{1}{p^0} \,
p^\mu \,
   \varphi_{0p}^\alpha \, \Big\{ f^{\mathrm{eq}}_p - \big[
   A^{\mbox{-}1} \, \bar{Q} \, F \big]_p \Big\}=
 J^{\mu\alpha}_{\mathrm{0}} + \delta J^{\mu\alpha},
 \end{eqnarray}
 where
 $J^{\mu\alpha}_{\mathrm{0}} \equiv \sum_{p} \, \frac{1}{p^0} \, p^\mu
 \, \varphi_{0p}^\alpha \, f^{\mathrm{eq}}_p$
 and
 $\delta J^{\mu\alpha} \equiv - \sum_{p} \, \frac{1}{p^0} \, p^\mu \,
 \varphi_{0p}^\alpha \, \big[A^{\mbox{-}1} \, \bar{Q} \, F \big]_p$.
Here $J^{\mu\alpha}_{\mathrm{0}}$ represents
 the currents of perfect-fluid part,
while $\delta J^{\mu\alpha}$
the  dissipative part.
 In deriving Eq.(\ref{eq:017}), we have used the following
 relation :
$\sum_{p} \, \frac{1}{p^0} \, (p \cdot \Ren{a}_p) \, \varphi_{0p}^{\alpha}
\, \big[ \bar{P} \, \psi \big]_p
= \sum_{p} \, \frac{1}{p^0} \, (p \cdot \Ren{a}_p) \, \varphi_{0p}^{\alpha}
\, \psi_p$,
obtained from
the definition $\bar{P} = f^{\mathrm{eq}} \, P \, f^{\mathrm{eq}\mbox{-}1}$
 and Eq.(\ref{eq:010}).
 Notice that the derivation of Eq.(\ref{eq:017}) is
 accomplished for arbitrary $\Ren{a}_p^\mu$ which is now dependent
on $\tau$ as well as on $\sigma$; in other words, we
 have a set of relativistic hydrodynamic equations for
viscous fluids, which have  still the freedom of choice
of $\Ren{a}_p^\mu$.
The is also one of the main results in the present work.


In the rest of the paper,
we shall show how  known relativistic dissipative hydrodynamic
equations are derived with a choice of the macroscopic frame vector
$\Ren{a}_p^\mu$.

As a simple but nontrivial choice, let us take the following set of
the macroscopic frame vectors with $\theta$ being a constant;
$\Ren{a}_p^\mu = (p\cdot u)^{\mbox{-}1} \, \{(p\cdot u) \, \cos\theta + m \,
\sin\theta \} \, u^{\mu} \equiv \theta_p^{\mu}$,
which leads to the following relations
$\Ren{\Delta}_p^{\mu\nu} = g^{\mu\nu} - u^\mu \, u^\nu \equiv
\Delta^{\mu\nu}$;
then one finds that
$(p \cdot u)^{\mbox{-}1} \, \{(p\cdot u) \, \cos\theta + m \, \sin\theta \}
\, \frac{\partial}{\partial \tau} = u^\mu \, \partial_\mu \equiv D$
and
$\Ren{\nabla}^\mu = \Delta^{\mu\nu} \, \partial_\nu \equiv \nabla^\mu$.
We notice that the factor $m$ is introduced simply to
make the expression dimensionless, so our method is also applicable to
the case of massless particles.

In the reduction of the currents in terms of the
fluid dynamical quantities,
the central task is the evaluation of
$A^{\mbox{-}1} \, \bar{Q} \, F$.
A straightforward but tedious manipulation leads to the following formula,
\begin{eqnarray}
  \big[ A^{\mbox{-}1} \, \bar{Q} \, F \big]_p =
  f^{\mathrm{eq}}_p \, \sum_{q} \, \mathcal{L}^{\mbox{-}1}_{pq} \,
\frac{1}{T} \,
  \Big( \Pi_q \, X_\theta + Q^\mu_q \, X_{\theta\mu} + \Pi^{\mu\nu}_q \,
X_{\mu\nu} \Big)\label{eq:099},
\end{eqnarray}
where $\mathcal{L}_{pq} \equiv (p \cdot \theta_p) \, L_{pq}$ being
independent
of $\theta^\mu_p$.
Here we have introduced the following quantities:
the dissipative currents, i.e.,
the bulk pressure,
$\Pi_p \equiv (4/3 -  \gamma) \, (p \cdot u)^2 + [(\gamma - 1) \, T \,
\hat{h} - \gamma \, T] \, (p \cdot u) - 1/3 \, m^2$,
the heat flow,
$Q^\mu_p \equiv - [(p \cdot u) - T \, \hat{h}] \, \Delta^{\mu\nu} \, p_\nu$
and the stress tensor,
$\Pi^{\mu\nu}_p \equiv 1/2 \, (\Delta^{\mu\rho} \, \Delta^{\nu\sigma} +
\Delta^{\mu\sigma} \, \Delta^{\nu\rho}
- 2/3 \, \Delta^{\mu\nu} \, \Delta^{\rho\sigma}) \, p_\rho \, p_\sigma$.
The coefficient of each dissipative current is
the respective dissipative thermodynamic force, i.e.,
$X_\theta \equiv (- z^2 \, \cos 2 \, \theta) \, \nabla \cdot u /
\{z^2 \, \cos^2\theta + z^2 \, (3 \, \gamma - 4) \, \sin^2\theta - 3 \, z \,
[1 - (\hat{h} - 1) \, (\gamma - 1)] \, \cos\theta \, \sin\theta\}$,
$X_{\theta\mu} \equiv \nabla_\mu \ln T - \cos\theta \, \nabla_\mu \ln (n \,
T) / (\hat{h} \, \cos\theta + z \, \sin\theta)$,
and
$X_{\mu\nu} \equiv 1/2 \, (\Delta_{\mu\rho} \, \Delta_{\nu\sigma} +
\Delta_{\mu\sigma} \, \Delta_{\nu\rho}
- 2/3 \, \Delta_{\mu\nu} \, \Delta_{\rho\sigma}) \, \nabla^\rho u^\sigma$,
where the following definitions have been used;
the reduced mass $z \equiv \frac{m}{T}$,
the particle density $n \equiv (2 \, \pi)^{\mbox{-}3} \, 4 \, \pi \, m^3 \,
\mathrm{e}^{\frac{\mu}{T}} \, \{ z^{\mbox{-}1} \, K_2(z) \}$,
the energy density $\epsilon \equiv m \, n \, \{ K_3(z) / K_2(z) -
z^{\mbox{-}1} \}$,
the reduced enthalpy per particle $\hat{h} \equiv \frac{\epsilon + n \, T}{n
\, T}$
and the ratio of the heat capacities $\gamma \equiv 1 + \{ z^2 + 3 \,
\hat{h} - (\hat{h} - 1)^2 \}^{\mbox{-}1}$,
with the modified Bessel functions $K_2(z)$ and $K_3(z)$.

Since the dissipative part of the currents has been obtained
for arbitrary $\theta$,
we can obtain the microscopic formulae for the transport coefficients, i.e.,
the bulk viscosity $\zeta$, the heat conductivity $\lambda$ and the shear
viscosity $\eta$,
which are nicely given in forms of the Kubo formula as follows:
Using the new inner product
${\langle \, \varphi \,,\, \psi \, \rangle}_\mathrm{eq} \equiv \sum_p \,
\frac{1}{p^0} \, f^\mathrm{eq}_p \, \varphi_p \, \psi_p$
and the evolved microscopic dissipative currents
$\big[ \Pi(s) \big]_p \equiv \sum_{q} \, \big[ \mathrm{e}^{s \, \mathcal{L}}
\big]_{pq} \, \Pi_q$ and so on,
we have
$\zeta \equiv (1/T) \, \int_0^\infty \!\! \mathrm{d}s {\langle \, \Pi (0)
\,,\, \Pi (s) \, \rangle}_\mathrm{eq}$,
$\lambda \equiv -(1/3T^2) \, \int_0^\infty \!\! \mathrm{d}s {\langle \,
Q^\mu(0) \,,\, Q_{\mu}(s) \, \rangle}_\mathrm{eq}$
and
$\eta \equiv (1/10T) \, \int_0^\infty \!\! \mathrm{d}s {\langle \,
\Pi^{\mu\nu}(0) \,,\, \Pi_{\mu\nu}(s) \, \rangle}_\mathrm{eq}$.
It is noteworthy that these transport coefficients are independent of
$\theta$
while the dissipative thermodynamic forces are dependent on it.

Although we can now write down the generic form of the currents
$J^{\mu\alpha}_{\mathrm{hydro}}$ for arbitrary $\theta$,
we shall give here them only for a few values of $\theta$ for the sake of
space,
but show that our generic currents included the known currents for
the relativistic fluid dynamics for a viscous fluid.

(A)~ With the simplest choice $\Ren{a}^\mu_p = u^\mu$, i.e., $\theta=0$, we
have
\begin{eqnarray}
  J^{\mu\alpha}_{\mathrm{hydro}} = \left\{
    \begin{array}{ll}
      \displaystyle{
        \epsilon \, u^\mu \, u^\nu - (p + \zeta \, X) \, \Delta^{\mu\nu} + 2
\, \eta \, X^{\mu\nu}
      } &
      \displaystyle{\alpha = \nu,}\\[2mm]
      \displaystyle{
        m \, n \, u^\mu - \frac{m}{\hat{h}} \, \lambda \, X^\mu
      } &
      \displaystyle{\alpha = 4,}
    \end{array}
  \right. \label{eq:039}
\end{eqnarray}
where $p \equiv n \, T$, $X \equiv - \nabla \cdot u$ and
$X^\mu \equiv \nabla^\mu \ln T - \hat{h}^{\mbox{-}1} \, \nabla^\mu \ln (n \,
T)$.
If we employ the constitutive equations;
$\Pi = \zeta \, X$, $Q^{\mu} = T \, \lambda \, X^{\mu}$ and
$\Pi^{\mu\nu} = 2 \, \eta \, X^{\mu\nu} - \Pi \, \Delta^{\mu\nu}$, it is
possible to
rewrite the equation in terms of the flows rather than the
thermodynamic forces as
\begin{eqnarray}
  \label{eq:039_a}
   J^{\mu\alpha}_{\mathrm{hydro}} =
   \left\{
    \begin{array}{ll}
     \displaystyle{\epsilon \, u^\mu \, u^\nu - p \, \Delta^{\mu\nu}
      + \Pi^{\mu\nu}} &
     \displaystyle{\alpha = \nu,}\\[2mm]
     \displaystyle{m \, n \, u^\mu - \frac{m}{\hat{h}T} \,Q^\mu} &
     \displaystyle{\alpha = 4.}
    \end{array}
   \right.
\end{eqnarray}
One should notice that
$J^{\mu\alpha}_{\mathrm{hydro}}$ in Eq.(\ref{eq:039}) or Eq.(\ref{eq:039_a})
agrees completely with the Landau theory for non-ideal fluids.
This was actually anticipated:
In fact, if we take a natural choice that $\Ren{a}_p^\mu = u^\mu$,
the physical quantity transported per unit$\tau$ becomes the microscopic
thermal energy $(p \cdot u)$,
which is reduced to the familiar form $m + \frac{m}{2} \, \vert
\frac{\Vec{p}}{m} - \Vec{u} \vert^2$
in the non-relativistic limit.
Thus it is natural that the currents $J^{\mu\alpha}_{\mathrm{hydro}}$
with the choice  $\Ren{a}_p^\mu = u^\mu$ becomes those in the  Landau frame.

(B)~ Another simple choice for the macroscopic frame vector is given with
$\theta=\pi/2$;
$\Ren{a}_p^\mu = \frac{m}{p \cdot u} \, u^\mu$.
A similar manipulation as the above gives the following currents
\begin{eqnarray}
  J^{\mu\alpha}_{\mathrm{hydro}} = \left\{
    \begin{array}{ll}
      \displaystyle{
        (\epsilon + 3 \, \zeta \, \tilde{X}) \, u^\mu \, u^\nu - (p + \zeta
\, \tilde{X}) \, \Delta^{\mu\nu}
        + \lambda \, T \, u^\mu \, \tilde{X}^\nu + \lambda \, T \, u^\nu \,
\tilde{X}^\mu
        + 2 \, \eta \, X^{\mu\nu}
      } &
      \displaystyle{\alpha = \nu,}\\[2mm]
      \displaystyle{m \, n \, u^\mu} &
      \displaystyle{\alpha = 4,}
    \end{array}
  \right.\label{eq:stuart}
\end{eqnarray}
where $\tilde{X} \equiv - \{1/3 \, (4/3 -  \gamma)^{\mbox{-}1}\}^2 \, \nabla
\cdot u$ and
$\tilde{X}^\mu \equiv \nabla^\mu \ln T$.
In terms of the flows, we have
\begin{eqnarray}
  J^{\mu\alpha}_{\mathrm{hydro}} = \left\{
    \begin{array}{ll}
      \displaystyle{
        \epsilon \, u^\mu \, u^\nu - p \, \Delta^{\mu\nu}
        + 3 \, \tilde{\Pi} \, u^\mu \, u^\nu + u^\mu \, \tilde{Q}^\nu +
u^\nu \, \tilde{Q}^\mu + \tilde{\Pi}^{\mu\nu}
      } &
      \displaystyle{\alpha = \nu,}\\[2mm]
      \displaystyle{m \, n \, u^\mu} &
      \displaystyle{\alpha = 4,}
    \end{array}
  \right.\label{eq:stuart_a}
\end{eqnarray}
where $\tilde{\Pi} = \zeta \, \tilde{X}$,
$\tilde{Q}^\mu = \lambda \, T \, \tilde{X}^\mu$
and $\tilde{\Pi}^{\mu\nu} = 2 \, \eta \, X^{\mu\nu} - \tilde{\Pi} \,
\Delta^{\mu\nu}$.
Since the dissipative term with the heat conductivity is
absent in $J^{\mu 4}_{\mathrm{hydro}}$,
one might be tempted to
identify the above equation with the Eckart one,
but it is not the case:
A detailed examination
tells us that Eq.(\ref{eq:stuart}) is
rather a modified or corrected version, we would say,  of
Stewart's\cite{mic003}.
This observation is
based on the fact
that Eq.(\ref{eq:stuart}) meets the Stewart's constraints, (v), (ii) and
(iii)
but not the Eckart's, (i), (ii) and (iii), mentioned
in the Introduction. The reason why we call Eq.(\ref{eq:stuart}) the
corrected version of Stewart equation will be
shortly explained.
A noteworthy point is the following:
In the Eckart theory
\footnote{In the original paper \cite{hen001}, basically the bulk viscosity
is not
 taken into account.}
used conventionally,
the appearance of the terms with the bulk viscosity is the same as the
Landau theory.
In contrast, Eq.(\ref{eq:stuart}) shows the different appearance
so that the bulk viscosity has an influence on the terms of
not only the pressure $p$ but also the energy density $\epsilon$.
We can show that our equation (\ref{eq:stuart}) satisfies the
second law of thermodynamics as will be explicitly shown
in a separate paper \cite{extend,extend2},
although the original Stewart equation may not.

(C)~ The other interesting choice of $\theta$ is $\theta = - \pi / 4$,
for which we have $\Ren{a}_p^\mu = \frac{(p \cdot u) - m}{\sqrt{2} \, p
\cdot u} \, u^\mu$, and
\begin{eqnarray}
  J^{\mu\alpha}_{\mathrm{hydro}} = \left\{
    \begin{array}{ll}
      \displaystyle{
        \epsilon \, u^\mu \, u^\nu - p \, \Delta^{\mu\nu}
        - \lambda \, T \, u^\mu \, \bar{X}^\nu - \lambda \, T \, u^\nu \,
\bar{X}^\mu
        + 2 \, \eta \, X^{\mu\nu}
      } &
      \displaystyle{\alpha = \nu,}\\[2mm]
      \displaystyle{
        m \, n \, u^\mu - \lambda \, T \, \bar{X}^\mu
      } &
      \displaystyle{\alpha = 4,}
    \end{array}
  \right. \label{eq:a039}
\end{eqnarray}
where
$\bar{X}^\mu \equiv z (\hat{h} - z)^{\mbox{-}1} \nabla^\mu \ln T - z
(\hat{h} - z)^{\mbox{-}2} \nabla^\mu \ln (n \, T)$.
The alternative expression in terms of the flows is
\begin{eqnarray}
  J^{\mu\alpha}_{\mathrm{hydro}} = \left\{
    \begin{array}{ll}
      \displaystyle{
        \epsilon \, u^\mu \, u^\nu - p \, \Delta^{\mu\nu}
        - u^\mu \, \bar{Q}^\nu - u^\nu \, \bar{Q}^\mu + \bar{\Pi}^{\mu\nu}
      } &
      \displaystyle{\alpha = \nu,}\\[2mm]
      \displaystyle{m \, n \, u^\mu - \bar{Q}^\mu} &
      \displaystyle{\alpha = 4,}
    \end{array}
  \right.\label{eq:a039_a}
\end{eqnarray}
where $\bar{Q}^\mu = \lambda \, T \, \bar{X}^\mu$ and $\bar{\Pi}^{\mu\nu} =
2 \, \eta \, X^{\mu\nu}$.
 In this frame, $\bar{X} \equiv X_{\theta = -\pi/4} = 0$ so that
 the associated bulk pressure $\bar{\Pi}( = \zeta \bar{X})$ disappears.
The unique feature of this equation
is the absence of terms with the bulk viscosity.
As far as we know,  this type of the dissipative fluid dynamical
equation is written down for the first time,
which was made possible by the introduction of the macroscopic
frame vector in the  powerful RG method.


Now we have seen that the fluid dynamical equation for dissipative
fluids in
the Eckart frame does not take the
form of Eckart, if we start from the underlying Boltzmann equation.
This fact is actually known for some time \cite{vanweert}.
 Nevertheless it is instructive
to see that
there can not exist the frame vector $\Ren{a}^\mu_p$
that  leads to the Eckart's constraints, (i), (ii) and (iii),
because it would help to elucidate the physical meaning of the
P- and Q-space in our formalism and their relation
to the phenomenological constraints imposed to the
dissipative part of the energy-momentum tensor $\delta J^{\mu\nu}$
and
the particle-number current $\delta J^{\mu 4}$, which are
defined just below Eq.(\ref{eq:017}).
Eq.(\ref{eq:099}) tells us that the dissipative part of
the one-particle distribution function $f_p$,
i.e., $- \big[ A^{\mbox{-}1} \, \bar{Q} \, F \big]_p$ can be
written as $f^\mathrm{eq}_p \, \phi_p$
with
$\phi_p \equiv - \big[ L^{\mbox{-}1} \, Q \, f^{\mathrm{eq}\mbox{-}1} \, F
\big]_p$.
Since the Q-space vector $\phi_p$ is orthogonal to
the five P-space vectors $\varphi^\alpha_{0p}$ by definition, we have
\begin{eqnarray}
\langle \, \varphi^\alpha_0 \,,\, \phi  \, \rangle = 0 \,\,\, \mbox{for}
\,\,\, \alpha = 0 ,\, 1 ,\, 2 ,\, 3 ,\, 4\label{eq:add005}.
\end{eqnarray}
Here the inner product is defined by Eq.(\ref{eq:inner}) where
$\Ren{a}^\mu_p$ enters with the form $(p \cdot \Ren{a}_p)$.
The notable point is that
these five identical equations exactly
correspond to the constraints on $\delta J^{\mu\nu}$ and
$\delta J^{\mu4}$.
In fact,
 Eq.(\ref{eq:add005}) with
the choice of (A) $\Ren{a}^\mu_p = u^\mu$ gives
$\sum_p \, \frac{1}{p^0} \, (p \cdot u) \, f^\mathrm{eq}_p \,
\varphi^\alpha_{0p} \, \phi_p = 0$,
which is equivalent to the set of equations
$u_\nu \, \delta J^{\mu\nu} = 0$ and
$u_\mu \, \delta J^{\mu 4} = 0$, because
$\delta J^{\mu\alpha} = \sum_p \, \frac{1}{p^0} \, p^\mu \,
\varphi^\alpha_{0p} \, f^\mathrm{eq}_p \, \phi_p$.
The former equation
may be re-expressed by
$u_\mu \, u_\nu \, \delta J^{\mu\nu} = 0$ and
$\Delta_{\mu\rho} \, u_\nu \, \delta J^{\mu\nu} = 0$.
Thus one can readily see
that these equations coincide with the Landau's constraints,
(i), (ii) and (iv).
Similarly,
with the choice of (B) $\Ren{a}^\mu_p = \frac{m}{p \cdot u} \, u^\mu$, we
have
$\sum_p \, \frac{1}{p^0} \, m \, f^\mathrm{eq}_p \, \varphi^\alpha_{0p} \,
\phi_p = 0$,
which means that
$\delta J^{\mu 4} = 0$
and $\delta J^\mu_{\,\,\,\mu} = 0$; the former equation is
equivalent to the set of equations
$u_\mu \, \delta J^{\mu 4} = 0$ and $\Delta_{\mu\nu} \, \delta J^{\mu 4} =
0$.
Thus one sees that
the Stewart's constraints, (v), (ii) and (iii) are derived.
Here we have used the on-shell condition $m^2 = p^\mu \, p_\mu$.
Now it is easy to show
 that there exists no
$\Ren{a}^\mu_p$ leading to the Eckart's constraints, (i), (ii) and (iii),
simultaneously.
In order to lead to the constraints (ii) and (iii) on $\delta J^{\mu 4}$,
Eq.(\ref{eq:add005}) with $\alpha = \mu$
requires that $(p \cdot \Ren{a}_p) = \mbox{const.}$, i.e.,
 independent of $p^\mu$.
On the other hand,
in order to lead to the constraint (i) on $\delta J^{\mu\nu}$,
Eq.(\ref{eq:add005}) with $\alpha = 4$
must lead to $(p \cdot \Ren{a}_p) = \mbox{const.} \, \times \, (p \cdot
u)^2$,
which is in contradiction with the condition for (ii) and (iii).
Thus
we can conclude that there exists
no $\Ren{a}^\mu_p$ leading to
 the Eckart's constraints (i), (ii) and (iii) simultaneously.

What we have seen is that
whenever the Eckart frame is taken
where the particle flow is constructed so as to have no
dissipative part,
the constraint on the dissipative part
of the energy-momentum tensor must satisfy (v) but not (i).
This fact suggests that the Eckart's theory
may not be realized as the slow dynamics
of the underlying Boltzmann equation, and hence nor have
microscopic foundation \cite{vanweert}.

One of the unique and important feature of the
hydrodynamical equations obtained in the present work
lies in the fact
that
 the thermal force driving the heat flow
contains $\nabla^\mu T$ but {\em no} $D u^\mu$ as in the
Landau equation,
which in contrast with the Stewart and Eckart equations.
We shall briefly argue that
the $D u^\mu$ terms should not exist on the general ground
on the basis of a natural assumption on the origin of the dissipation;
the detailed argument will be presented
in  separate papers \cite{extend2}.

First of all,
we can trace back the absence of $D u^\mu$ terms
in the thermal forces
to our ansatz that the dissipation comes solely from the spatial
inhomogeneity at the rest frame.
This ansatz is clearly expressed in  the choice of $\Ren{a}^\mu_p$:
Although the generic form of $\Ren{a}^\mu_p$ which
keeps the Lorentz covariance can be given as
$\Ren{a}^\mu_p = \alpha(p \cdot u) \, u^\mu + \beta(p \cdot u)
 \, \Delta^{\mu\nu} \, p_\nu$,
where $\alpha(x)$ and $\beta(x)$ are the arbitrary Lorentz-scalar functions,
we have intentionally
put $\beta(x) = 0$ so that
the differential operator $\Ren{\nabla}^\mu$ in the
last term in Eq.(\ref{eq:004}), which
gives rise to the deviation from the local equilibrium
 and hence the dissipation,
agrees with the spatial derivative, i.e., $\Ren{\nabla}^\mu = \nabla^\mu$.
As mentioned before,
the non-relativistic Navier-Stokes equation can be nicely
derived by the same ansatz that the spatial inhomogeneity
leads to the dissipation \cite{env007,env008}, and 
the present approach is simply a covariantization of the non-relativistic
case.
On the other hand, the ansatz used by Landau or Stewart is that
the dissipation comes from both of the spatial and the temporal
inhomogeneity
at the rest frame, although no $D u^\mu$ terms appear in the Landau
equation.

In a separate paper,
we shall show that
the covariant dissipative hydrodynamic equations
constructed  in the phenomenological
way as was done by Eckart, Landau, Stewart so on but
based on the same ansatz as that in the present work,
completely agree
with the equations derived in the present RG method
for the Landau and Eckart frame, i.e., with the two choices
for $\Ren{a}^\mu_p$.


In summary, we have derived generic covariant hydrodynamical
equations for viscous fluids as a reduction of dynamics
from relativistic Boltzmann equation in a mechanical way with no heuristic
arguments on the basis of  so called the renormalization-group method:
This was made possible by introducing the macroscopic frame vector which
defines the macroscopic Lorenz frame in which the slow dynamics
is described.
Our generic equation  includes 
not only the Landau equation\cite{hen002}
but also  a modified Eckart/Stewart equation\cite{mic003,cau001} 
and a novel one which does not contain the bulk viscosity term.

We notice that the derived equation is consistent with the underlying
kinetic equation, so the equations and also the method developed here
may be useful for the analysis of the system where
 the proper dynamics describing the system changes from the
hydrodynamic to kinetic regime, as in the system near the freeze-out region
in the RHIC experiment.

Our method is so mechanical and simple that
it is successfully applicable
to derive the dissipative hydrodynamics for the multi-component system
in Landau frame \cite{extend}.
In the present work,
the fluid dynamical variables
correspond to the conserved quantities in the collision process
and span the five-dimensional P space, which is an
invariant manifold in the terminology of the dynamical systems.
One may suspect that the method could apply to derive the
Israel-Stewart equation \cite{mic004} with fourteen variables in accordance
with Grad's moment method \cite{grad}.
The answer is affirmative: We have found \cite{extend}
that a simple  extension of the
P-space in our formalism gives the equation equivalent to Grad's
fourteen-moment method \cite{grad}.
These extensions  will be reported elsewhere.
In passing, we emphasize that our method itself has a universal
nature and can be applied to derive
a slow dynamics from other kinetic equations than the simple Boltzmann
equation.

We thank Berndt Muller and Tetsufumi Hirano for their interest
in this work and encouragement.
T.K. is supported by Grant-in-Aid
for Scientific Research by Monbu-Kagakusyo(No. 17540250).

\end{document}